**Title**

Detergents and chaotropes for protein solubilization before two-dimensional electrophoresis

**Authors**

Thierry RABILLOUD*, Sylvie LUCHE*, Véronique SANTONI†, Mireille CHEVALLET*
* : Laboratoire, d'immunologie, DRDC/ICH, INSERM U 548, CEA Grenoble, 17, rue des martyrs, F38054 GRENOBLE CEDEX 9 FRANCE
†: Biochimie et Physiologie Moléculaire des Plantes, Agro-M/INRA/CNRS/UM2 UMR 5004, 2 place Viala, F-34060 Montpellier cedex, France

**Abbreviations**
ASB14: 3-(tetradecanoylamidopropyl dimethylammonio) propane 1-sulfonate
C7BzO: 3-(4-Heptyl)phenyl-3-hydroxypropyl)dimethylammoniopropanesulfonate
C13E10 deca(ethylene oxide) tridecyl ether
CTAB: cetyl trimethyl ammonium bromide
DTAB: dodecyl trimethyl ammonium bromide
DTT: dithiothreitol
CA-IEF: carrier ampholyte isoelectric focusing
IEF: isoelectrofocalisation
IPG : immobilized pH gradient
p*I*: isoelectric point
TBP : tri-butyl phosphine
TCEP: tris-carboxyethyl phosphine
SDS-PAGE: sodium dodecyl sulfate-polyacrylamide gel electrophoresis
2D: two-dimensional

**Abstract**
Because of the outstanding separating capabilities of two-dimensional electrophoresis for complete proteins, it would be advantageous to be able to apply it to all types of proteins. Unfortunately, severe solubility problems hamper the analysis of many classes of proteins, but especially membrane proteins. These problems arising mainly in the extraction and isoelectric focusing steps, solutions are sought to improve protein solubility under the conditions prevailing during isoelectric focusing. These solutions deal mainly with chaotropes and new detergents, which are both able to enhance protein solubility. The input of these compounds in proteomics analysis of membrane proteins is discussed, and future directions are also discussed

**Key words**
Hydrophobic protein. Membrane proteins. Protein solubilization. Isoelectric focusing. Chaotropes. Detergents. Zwitterionic detergents.

# 1. Introduction

The solubilization of proteins for 2D electrophoresis-based proteomics is a difficult task. As a matter of facts, it is almost one of the worst possible setups for protein solubilization. For example, proteins must reach their pI, which is also the minimum of solubility, and still stay soluble at the pI. Moreover, as the protein mobility decreases when proteins get close to their pI, IEF is performed under strong electric fields (200V/cm is not uncommon compared to the 10-20V/cm used for SDS-PAGE). This means in turn that salts and ionic compounds in general are almost banned in IEF. Moreover, any solubilizing agent used prior to IEF must not change the original pI of the proteins. Consequently, this precludes the use of strong ionic detergents such as SDS. However, low amounts (up to 0.03% w/v) of ionic detergents can be used, provided that conditions favoring the exchange of SDS for other, non-ionic detergents are used in IEF [1-3]. This ensures removal of bound SDS from the proteins, but this also means that the benefits of SDS are lost for the IEF dimension. However, the use of SDS has been often recommended as a way to ensure a complete initial solubilization before IEF. Apart from these problems arising from the protein world, other problems are frequently encountered in many biological samples. These problems arise form the non-proteinaceous compounds that can be present in the sample. A canonical example is the one of nucleic acids, which completely blur the 2D electrophoresis pattern when present at too high a concentration [4]. Nucleic acids act as mobile ion exchangers at the low ionic strength required by IEF, thereby creating severe artefacts. Other classes of compounds (lipids, salts etc.) can be encountered in many samples and create their artefacts. This is especially true for plant-derived samples, due to the ability of plant tissues to synthetize a host of compounds with varied structures. For example, phenolics and chlorophyll can be very abundant in some plant tissues and completely ruin protein separation in proteomics techniques.

There are thus different problems depending on the starting material. When starting from total tissues, the main problem is generally the interference arising from non-proteinaceous compounds. These aspects are described in other chapters from this book. This chapter will therefore focus on the other side of the problem; i.e. the intrinsic solubilization of proteins, and will mainly deal with the chaotropes and detergents used for initial solubilization and for IEF.

As mentioned earlier, the constraints present in IEF limit the choice to chemicals showing no net electric charge in solution over the pH range used for IEF, i.e. to nonionic or zwitterionic compounds. This narrows the choice for chaotropes to the amide and urea families, as guanidines and amidines are charged below pH 12. Among the possible chaotropes, urea has been used for quite a long time. More recently, the addition of thiourea to urea as an additional chaotrope has shown interesting features for protein solubilization [5] but also to limit protease action [6]. The role of chaotropes in the solubilization process is to break the non-covalent interactions between the various molecules present in the sample (e.g. hydrogen bonds, dipole-dipole interactions, ad hydrophobic interactions), and to unfold the proteins. Although ionic bonds are not directly affected by nonionic chaotropes such as urea and thiourea, the influence of these chaotropes on the dielectric constant of water also alters the strength of the ionic bonds.

On the detergent side, it is quite cleat that the uncharged detergents are clearly much less efficient than the ionic ones. Ionic detergents make a charged "coat" on the protein molecules, so that the protein-detergent complexes repel each other via ionic interactions, thereby preventing protein aggregation. Unfortunately, ionic detergents cannot be used for IEF-based 2D separation. However, they can be used in methods using differential zone electrophoresis, and a short example will be given in this chapter.

Among the wide choice of commercially available uncharged detergents, two subfamilies can be distinguished. Nonionic detergents do not have any charges on the molecule, while zwitterionic detergents have an equal number of negative and positive charges on their molecules. Depending on the pK of the ionizable groups on the molecules, some detergents can be ionic in a certain pH range (where at least one group is titrated) and zwitterionic in another pH range, while other detergents can be zwitterionic on the complete pH range. As an example of the two cases, classical betaines (bearing a quaternary ammonium and a carboxylic group) are positively charged at low pH, when the carboxylic group is not fully deprotonated. When the carboxylic group is fully deprotonated, i.e. more than 2 pH units above the pK, they behave as a zwitterionic detergent. Oppositely, sulfobetaines, bearing a quaternary ammonium and a sulfonic group, are zwitterionic over the 0-14 pH range, as both groups are ionized in this range. As a matter of facts, only detergents completely zwitterionic over the pH range of interest can be used for IEF.

The two "historical" detergents used for 2D electrophoresis are on the one side Triton X100 (or NP-40), a nonionic detergent and on the other side the zwitterionic detergent CHAPS. Both have been used extensively in combination with urea, and have not proved very efficient for the solubilization of sparingly soluble proteins, e.g. membrane proteins [7]. However, recent work has shown that either specially-designed zwitterionic detergents [8-10], or carefully-selected non ionic detergents [11] can solubilize membrane proteins. It is interesting to note that Triton X100 is poorly efficient when used with urea alone and much more efficient in urea-thiourea [11]. This observation extends to other detergents of the oligo ethylene glycol family, such as the Brij ® detergents, i.e. linear alkyl oligo ethylene glycol compounds [11]. However, the most efficient nonionic detergents belong to the glycoside family (e.g. octyl glucoside, dodecylmaltoside), and the latter seems to be efficient in urea alone [12] as well as in urea-thiourea [11, 13]. An example of the variations in protein solubilization induced by the choice of detergent can be seen in figure 1. The multiple variables playing a role in the solubilization process have also been investigated in [14]. However, it should not be concluded from the preceding that dodecyl maltoside is the absolute best choice for protein solubilization for 2D electrophoresis. The optimal detergent will depend on the sample. However, there is some kind of a shortlist, based on previous work. So the best candidates for protein solubilization, at least as a first screen, are to be chosen among dodecyl maltoside, ASB14, C7BzO, Brij56 and C13E10, Chaps being a good choice for soluble proteins

## 2. Materials

### *2.1. Biological material*

*Arabidopsis thaliana* membrane preparations are obtained according to Santoni *et al*. (this book).

## 2.2. Equipment

1. A tabletop ultracentrifuge, used for membrane preparation and to remove unsolubilized proteins
2. Immobilised pH gradient (IPG, linear and non-linear pH gradient from 3 to 10, 18 cm length, Amersham Pharmacia Biotech).
3. IPGphor apparatus: for isoelectrofocalisation of proteins (Amersham Pharmacia Biotech).
4. Tube gels electrophoresis Setup (BioRad), for first dimension gel electrophoresis
5. Protean II: for SDS-PAGE electrophoresis (Biorad).

## 2.3. Reagents

### 2.3.1. Products and stock solutions

1. Dodecyl maltoside, Triton X100, and CHAPS are best used from 20% (w/v) stock solutions in water. These solutions should be stored at 4°C and show limited conservation (a few weeks)
2. C13E10, Brij 56, ASB14 are best used from 20% (w/v) stock solutions in ethanol/water (50/50 v/v). These solutions are stable for months at room temperature
3. Cationic detergents (DTAB, CTAB, benzalkonium chloride) are used as a 20% (w/v) stock solution in water. these solutions are stable at room temperature but very sensitive to temperature. They sometimes need to be warmed at 37-40°C to redissolve the detergent prior to use.
4. Urea stock solution for IEF. It is difficult to go beyond 9M urea at room temperature, which is the concentration used when urea is the sole chaotrope. This means that urea is added as a solid (**Note 1**)
5. Urea-thiourea stock solution. The final chaotrope concentrations are 7M urea and 2M thiourea. This means that a 1,25x concentrated solution can be prepared, which is then simpler to use that reweighing small amounts of solid urea and thiourea for each sample. For 10 ml of this concentrated solution, weigh 5.25g of urea and 1.9g of thiourea. Some detergents (e.g. CHAPS or Triton X-100) which are fully compatible with urea, can be added at this stage. Other detergents, which show a more limited urea compatibility (e.g. ASB14), must be added only when the solution is diluted to the final strength. A total volume of 4.2 ml of liquid must be added to the urea and thiourea to make up for 10 ml (see also **Note 2** and **Note 3**). This solution is stable for months if stored frozen at -20°C.
6. urea solution for zone electrophoresis. As urea is used at 4M final concentration, it is quite convenient to prepare a 8M stock solution. For 10 ml, dissolve 4.8 g of urea in 6.4 ml of water. this solution is stable to 2-3 days at +4°C.
7. acidic solubilization buffer for zone electrophoresis. 1M potassium dihydrogen phosphate + 1μl/ml 85% phosphoric acid.
8. Tributylphosphine is a liquid (4M when pure). A 40 fold dilution in dimethyl formamide is made just prior to use. This solution is further diluted 50 fold in the sample solution.
9. Tris carboxyethyl phosphine is a solid. A 1M stock solution in water is made, and stable

for months at -20°C

# 3. Methods (see Note 4)

## 3.1 Solubilization in urea for IEF

### 3.1.1. Solubilization from a solid sample, e.g. tissue or cell pellet
In this case, the sample volume can often be neglected in the final solubilization volume. A sample solution containing urea (9 to 9.5 M final concentration, see Note 1), the selected detergent (taken in the list above in 2.3.1) at 2 to 4% w/v concentration, carrier ampholytes (0.4% w/v for IPG, 2% w/v for CA-IEF) and a reducing agent (50mM DTT or 5mM TBP or 5mM TCEP). This solution is added to the solid sample, resulting in a liquid extract. Protein extraction is helped by sonication in a water bath sonicator for ca. 30 minutes. Unsolubilized material is best removed by ultracentrifugation for 30 minutes at 200,000g at room temperature.

### 3.1.2. Solubilization from a suspension or solution
In this case, the volume of the sample must generally be taken into account. It is thus necessary to calculate the final solution. As a rule of thumb, the sample volume can represent up to 35% of the final extraction volume. Solid urea, water and stock solutions of the detergent, ampholytes and reducer are used in addition to the liquid sample to build the extraction solution.

## 3.2. Solubilization in urea-thiourea for IEF
With the spreading of IPG with sample application by in-gel rehydration [15], rather large sample volumes can be used. This is especially true when home made strips are used, as these can be made wider than commercial IPG strips and thus accommodate a larger volume (up to 1 ml). It is thus often possible to use a dilution approach with the concentrated chaotrope solution and the solid sample resuspended in a minimal volume of water or the liquid sample. If the detergent can be predissolved in the concentrated chaotrope solution, which can also contain the reducer, then the sample volume can represent up to 20% of the total extraction volume. If the detergent must be added only at last, with a urea concentration not exceeding 8M, then it is more convenient to use 1 volume of sample, 1 volume of detergent stock solution, and to add 8 volumes of concentrated chaotrope solution. If this approach leads to too high a volume, two alternate approaches can be considered:
i) introduce in a sample tube a volume of a stock detergent solution equal to the sample volume. Evaporate the solvent in a SpeedVac. Add the sample and 4 volumes of concentrated chaotrope solution.
ii) Consider the sample volume will make 40% of the final sample volume. Weigh the corresponding amounts of urea, thiourea and solid detergent. Dissolve with the sample in a bath sonicator.
In all cases, an extraction time of 30-60 minutes at room temperature is optimal before centrifugation (200,000g, 30 minutes, room temperature) to remove unsolubilized material.

## 3.3. Solubilization for zone electrophoresis

The solubilization with SDS is not considered in this section, which will deal only with the solubilization in urea and cationic detergents prior to off-diagonal electrophoresis [16-17]. The calculations for making the extraction solution are simple, since the sample represents 1/4 of the final extraction volume. To the liquid sample are added (in this order, and expressed as a fraction of the initial sample volume):
0.4 volume of 20% (w/v) cationic detergent stock solution
0.2 volume of reducer
0.4 volume of acidic phosphate buffer
2 volumes of 8M urea
The solution is extracted for 30 minutes in a bath sonicator. Centrifugation at 10,000g for 15 minutes at room temperature may be needed to remove precipitated material.

## 4. Notes

1. The partial specific volume of urea is a useful number to know to make concentrated urea solution. 1 g of urea occupies 0.75 ml in solution. In the same order for most detergents, 1g occupies 1 ml. This is also true for thiourea. Urea must also never be warmed above 37°C to limit protein carbamylation. As an example, 1 ml of aqueous extract is added to 900 mg urea. this results in a 1.675 ml of a 9M urea solution. Otherwise, 1g urea is added to 1 ml aqueous extract, resulting in 1.75 ml of a 9.5M urea solution.
2. As a matter of facts, most of these extraction solution contain less than 50% water. This means that dissolution of the solids is rather difficult to perform, especially because high temperatures cannot be used (see previous note). The use of a water bath sonicator (marketed for cleaning objects and glassware) is of great help for this difficult solubilizations.
3. many detergents are not fully compatible with urea. depending on the detergent structure, on the urea concentration and on the temperature, insoluble detergent-urea complexes can form. Detergents with linear alkyl chain are especially prone to this problem (e.g. ASB 14, Brij 56) which completely prevents the use of commercial linear sulfobetaines, which do not stand more than 4M urea.
4. Many detergents strongly interfere with some popular protein assay methods, while others are plagued by reducing compounds (see the corresponding chapter in this book). It is therefore recommended to take this dimension into account when performing protein solubilization. In some cases, one can end with a solubilization cocktail which is incompatible with any protein assay method. In this case, it is often advisable to determine the protein concentration in the initial sample, prior to extraction, especially if the sample is a suspension. This means in turn that it will not be possible to assess the efficiency of the solubilization process.

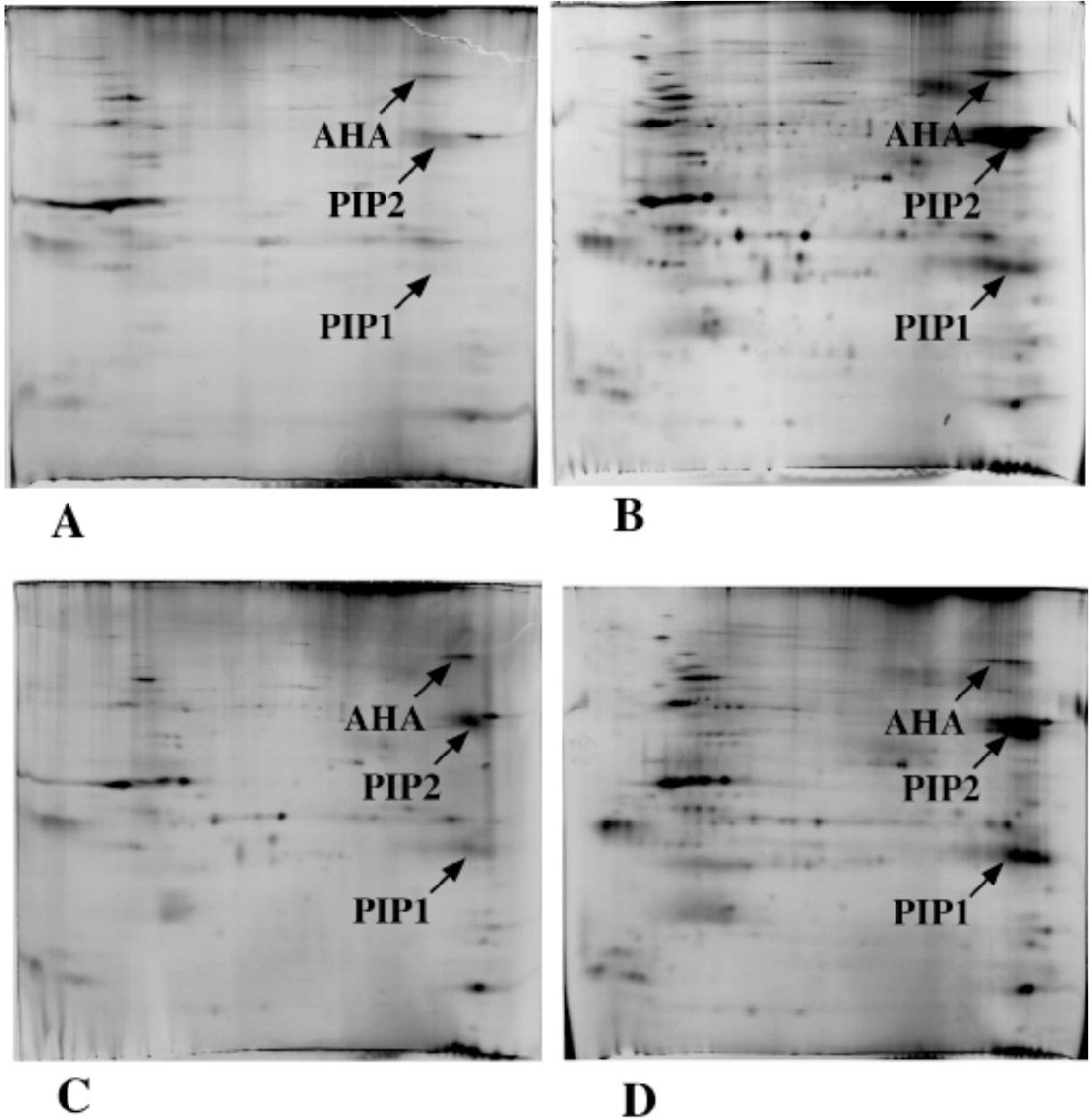

Figure 1. 2D electrophoretic separation of A. thaliana leaf plasma membrane proteins
60μg of leaf plasma membrane proteins were loaded on the 2D gels. The first dimension is a 3-10 linear pH gradient, and the second dimension a 10% acrylamide gel. H+ ATPase (AHA), aquaporin monomer (PIP1) and aquaporin dimer (PIP2) are indicated by arrows. The proteins are extracted and focused in a solution containing 7M urea, 2M thiourea, 20 mM DTT, 0.4% carrier ampholytes and A: 4% CHAPS; B: 2% dodecyl maltoside; C: 2% C7BzO; D: 2% ASB14